\documentclass[%draft,
preprint,aps,%superscriptaddress,
showpacs%,nofootinbib,eqsecnum,preprintnumbers,twocolumn,prl
]{revtex4}

\usepackage{amsmath}
\usepackage{amssymb}
\usepackage{graphicx}
\usepackage{bm}
\usepackage{epsfig}
\input{epsf}
\begin{document}

\preprint{Revised version}

\title{
Vortices and Superfields on a Graph
}

\author{Nahomi Kan}
\email{kan@yamaguchi-jc.ac.jp}
\affiliation{Yamaguchi Junior College, %1346-2 Daidou, 
Hofu%-shi
, Yamaguchi 747-1232, Japan}
\author{Koichiro Kobayashi}
\email{m004wa@yamaguchi-u.ac.jp}
\author{Kiyoshi Shiraishi}
\email{shiraish@sci.yamaguchi-u.ac.jp}
\affiliation{
Graduate School of Science and Engineering, Yamaguchi University
\\ %Yoshida, 
Yamaguchi%-shi
, Yamaguchi 753-8512, Japan
}

\begin{abstract}
%%%%%%%%%%%%%%%%%%%%%%%%%%%%%%%%%%%%%%%%%%%%%%%%%%%%%%%%%%%%%%%%%%%%%
We extend the dimensional
deconstruction by utilizing the knowledge of graph
theory. In the dimensional deconstruction, one uses the
moose diagram to exhibit the structure of the `theory space'. We
generalize the moose diagram to a general graph with oriented edges. 
In the present paper, we consider only the $U(1)$ gauge symmetry. 

We also
introduce supersymmetry into our model by use of superfields. We
suppose that vector superfields reside at the vertices and chiral
superfields at the edges of a given graph. Then we can consider
multi-vector, multi-Higgs models. In our model,
$[U(1)]^p$ (where $p$ is the number of vertices) is broken to a single
$U(1)$. Therefore for specific graphs, we get vortex-like classical
solutions in our model. We show some examples of the graphs admitting the
vortex solutions of simple structure as the Bogomolnyi solution.

%%%%%%%%%%%%%%%%%%%%%%%%%%%%%%%%%%%%%%%%%%%%%%%%%%%%%%%%%%%%%%%%%%%%%
\end{abstract}

\pacs{02.10.Ox, 11.10.Lm, 11.27.+d, 11.30.Qc
}
\maketitle

\section{Introduction}
%Motivated by Dimensional Deconstruction~\cite{Deconstruction} (DD),\\
%We consider multi-vector, multi-Higgs models and their Stueckelberg-type
%variation.
%For notation, please consult the reference~\cite{KSJMP}.
%DD---mass matrices, almost free theory. interaction?
%susy (and local symmetries) determines interactions.
%$U(1)$.

Recently, `Higgsless theories' are eagerly studied by many
authors~\cite{Higgsless,ThreeSite}. Most of these models are derived from
or related with the method of the dimensional deconstruction
(DD)~\cite{Deconstruction}, which leads to the breakdown of
electroweak symmetry.

%%%%%%%%

\begin{figure}[h]
\begin{center}
\includegraphics[width=15cm]
{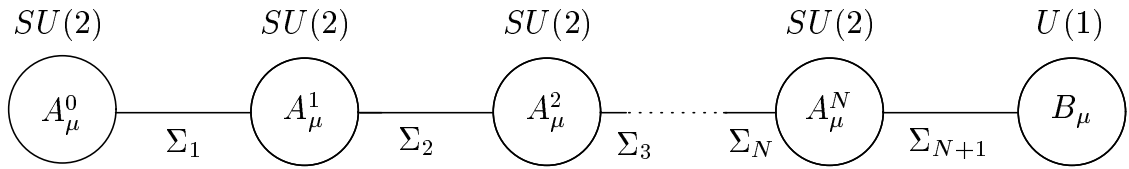}
\end{center}
\caption{A moose diagram.
There are $N+1$ $SU(2)$ gauge fields and a $U(1)$ gauge field.
Each gauge field exists on each site represented by a small circle.
The coupling constant of the gauge fields $A_{\mu}^i$ is $g_{i}$
($i=0,1,2,\cdots,N$), and the coupling constant of the $U(1)$ gauge field
$B_{\mu}$ is $g_{N+1}$. The vacuum expectation value of the scalar fields
$\Sigma_{i}$ is $f_{i}$ ($i=1,2,\cdots,N,N+1$).}
\label{moosediagram001}
\end{figure}

The typical structure of DD is shown diagrammatically in
FIG.~\ref{moosediagram001}. This model incorporates the $[SU(2)]^{N+1}
\otimes U(1)$ gauge group and $N+1$ nonlinear-sigma-model fields. If $N$
is equal to one, the number of the site is three in
FIG.~\ref{moosediagram001}. The three-site Higgsless
model~\cite{ThreeSite} is in this category.
In the generic scenario, the $[SU(2)]^{N+1} \otimes U(1)$ gauge group is
broken to $U(1)$.

The moose diagram like FIG.~\ref{moosediagram001} naturally leads to the
Lagrangian of the model.  This moose diagram indicates a relation between
gauge fields and scalar fields. We will generalize this relation
in the context of graph theory.
We can express the relation between gauge fields and scalar
fields in a graph, which is just a complex moose.  
We wish to call this theory based on a graph as `graph dimensional
deconstruction' (GDD). The idea of GDD has already been published as
Ref.~\cite{KSJMP}.

In the present work, we propose another idea of using superfields to
introduce supersymmetry (SUSY) into the model. We assign vector
superfields to vertices and chiral superfields to edges of a graph.  This
is another extension of the DD. 

In the beginning, both DD and SUSY are to provide the
mechanism of solving the gauge hierarchy problem.
The motivations of including SUSY are, nevertheless, claimed as
follows. First of all, we should think that every field theory has 
SUSY at very high energy, because the correct or controlled UV
behaviors are believed, or because of superstring theory or M-theory.
The second motivation comes from the necessity of more symmetries.
Because  DD and GDD are basically the mechanism of controlling the mass
spectrum of field theory, we need more symmetry to determine the 
(self-)interaction of fields. Thus we consider the supersymmetric
extension of the GDD model here.

In this paper, we consider only the Abelian theory.
For notation, please consult Ref.~\cite{KSJMP}.

%\newpage
\section{A review of field theory on a graph (or GDD)}
\begin{figure}[h]
\begin{center}
\includegraphics[height=2.5cm]
{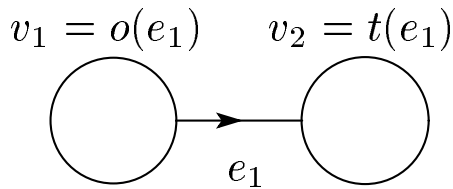}
\end{center}
\caption{
The simplest graph, constructed by two vertices and an edge.
A vertex ${v}_i$ is identified by $i$, where $i$ is a label for each 
vertex. In the same way, an edge $e_i$ is identified by $i$, where $i$
is a label for each edges. The arrow means a direction of the edge.
This edge is called an oriented edge.
In terms of the oriented edge, the original vertex $v_1$ is $v_1=o(e_1)$ 
and the terminal vertex $v_2$ is $v_2=t(e_1)$. This oriented graph
corresponds to the generalized moose diagram.}
\label{pathgraphsample}
\end{figure}

A graph $G(V,E)$ consists of a set of vertices $V$ and a
set of  edges $E$. A vertex is connected with another one by an edge.
We let the number of the vertices be $p$, $p\equiv\#V$ and the
number of the edges be $q$, $q\equiv\#E$.
In FIG.~\ref{pathgraphsample}, we show the simplest graph with $p=2$ and
$q=1$, constructed by two vertices and an edge.

We consider a simple Abelian theory.
Abelian gauge fields reside at vertices and scalar fields reside at edges.
The $U(1)$ transformation is defined at each vertex.
The Lagrangian density is 
\begin{equation}
{\cal L}=-\frac{1}{4}\sum_{v\in
V}F_{\mu\nu}^vF^{\mu\nu}_v-\sum_{e\in
E}({\cal D}_\mu U_e)^\dagger({\cal D}^\mu U_e)\, ,
\end{equation}
where the covariant derivative is
\begin{equation}
{\cal
D}^{\mu}U_e=(\partial^{\mu}+igA^{\mu}_{t(e)}-igA^{\mu}_{o(e)})U_e\,,
\end{equation}
with $|U_e|^2=f^2$.

If we rewrite $U_e$ as $U_e=f\,e^{ia_e}$, the real scalar fields $a_e$ 
act as the Stueckelberg fields.\cite{Stu} The number of physical massless
scalar fields is $q-p+1$, or the number of closed circuits involved in the
graph, because $p-1$ scalar degrees of freedom are absorbed by the to-be
massive vector fields. If and only if the graph is {\it tree} (or absent
from closed circuits), the scalar fields disappear from the physical
spectrum.

The $(mass)^2$ matrix of vector fields $M^2_A$ is given by
$2g^2f^2\Delta$, where the $(p, p)$ matrix
\begin{equation}
\Delta\equiv EE^T\,,
\end{equation}
is called as the graph Laplacian and the $(p,q)$ matrix $E$ is the
incidence matrix
\footnote{Unfortunately, the symbol $E$ is used for the incidence matrix
and for the set of edges. Please do not confuse them.}  defined as
\begin{equation}
(E)_{ve}=\left\{
\begin{array}{cc}
1 & {\rm if~}v=o(e)\\
-1 & {\rm if~}v=t(e)\\
0 & {\rm otherwise}
\end{array}
\right.\,.
\label{Ei}
\end{equation}
Here $v=o(e)$ means that the vertex $v$ is the origin of the edge $e$ and
$v=t(e)$  means that the vertex $v$ is the terminus of the edge $e$. 
The $(q, p)$ matrix $E^T$ is the transposed matrix of $E$.

For more general cases, one might consider individual coupling constants
for vertices as
\begin{equation}
{\cal
D}^{\mu}U_e= ( \partial^{\mu}+ig_{t(e)}A^{\mu}_{t(e)}-ig_{o(e)}A^{\mu}_{o(e)}) U_e\,,
\end{equation}
and $|U_e|^2=f_e^2$ for each edge. 
In this case the mass matrix becomes
\begin{equation}
M^2_A=2GEF^2E^TG=2(GEF)(GEF)^T,\end{equation}
where the diagonal matrices $G$ and $F$ are given by
\begin{equation}
(G)_{vv'}=\left\{
\begin{array}{cc}
g_v & {\rm if~}v=v' \\
0 & {\rm otherwise} 
\end{array}
\right.\,,\quad
(F)_{ee'}=\left\{
\begin{array}{cc}
f_e & {\rm if~}e=e' \\
0 & {\rm otherwise}
\end{array}
\right.\,,
\label{GF}
\end{equation}
respectively.

To summarize this section:
In the GDD model, the mass spectrum is given by eigenvalues of the graph
Laplacian or the related matrix constructed from the incidence matrix of
the graph.

%\newpage
\section{The use of the Stueckelberg superfield}
Next we incorporate SUSY into the GDD model.
We use superfields~\cite{SUSY} to this end.

In this paper, we consider that vector
superfields $\{V_v\}$ exist on vertices. We still impose the $U(1)$
transformation on $V_v$ at each vertex as
\begin{equation}
V_v\rightarrow V_v+i(\Lambda_v-\overline\Lambda_v)\,,
\end{equation}
where $\Lambda_v$ is a chiral superfield.
Then the invariant superfield is defined as usual:~\cite{SUSY}
\begin{equation}
W_\alpha^v=-\frac{1}{4}\overline{D}\,\overline{D}\,D_\alpha
{V_v}\, .
\end{equation}
The kinetic term of the vector field can be created from this for each
vertex.

Further we introduce a chiral superfield $S_e$ at each edge. The
superfield 
$S_e$ is assumed to be transformed as:
\begin{equation}
{S_e}\rightarrow S_e-i\Lambda_{t(e)}+i\Lambda_{o(e)}\,.
\end{equation}

%\begin{equation}
%e^{-S_e}\rightarrow e^{-2i\Lambda_{t(e)}}\,
%e^{-S_e}\,e^{2i\Lambda_{o(e)}}
%\end{equation}

Then we can write the Stueckelberg term~\cite{susystuckelberg}
\begin{equation}
(V_{t(e)}-V_{o(e)}+S_e+\overline S_e)^2\, ,
\end{equation}
and a gauge invariant term for the interaction with scalars
\begin{eqnarray}
{\cal L}&=&\sum_{v\in
V}\frac{1}{4g_v^2}\left(\left.W^\alpha_vW^v_{\alpha}\right|_{\theta\theta}
+\left.\overline{W}^v_{\dot{\alpha}}
\overline{W}^{\dot{\alpha}}_v\right|_{\bar\theta\bar\theta}\right)
+\sum_{e\in
E}\,\left.2f_e^2(V_{t(e)}-V_{o(e)}+S_e+\overline S_e)^2
\right|_{\theta\theta\bar\theta\bar\theta}.
\end{eqnarray}

The bosonic part of the theory is found to be
\begin{eqnarray}
{\cal L}_b&=&-\sum_{v\in
V}\frac{1}{4g_v^2}F_{\mu\nu}^vF^{\mu\nu}_v-\sum_{e\in E}
\frac{2f_e^2}{2}(A_{t(e)}^\mu-A_{o(e)}^\mu+\partial^\mu
a_e)^2-\frac{1}{2}\sum_{e\in
E}2f_e^2(\partial^\mu\rho_e)^2  \notag \\ 
&~&+\sum_{v\in V}\frac{1}{2g_v^2}D^2_v
+2\sum_{e\in E}2f_e^2|F_{Se}|^2
+\sum_{e\in
E}2f_e^2(D_{t(e)}-D_{o(e)})\rho_e\,,
\end{eqnarray}
where the notation of component field is rather standard one and is
gathered in Appendix~\ref{cSF}.

Eliminating the auxiliary fields $F_{Se}$ and rescaling $\rho_e$,
gauge fields and $D_v$ to have canonical kinetic terms we get
\begin{eqnarray}
{\cal L}_b&=&-\frac{1}{4}\sum_{v\in
V}F_{\mu\nu}^vF^{\mu\nu}_v-\sum_{e\in E}
\frac{2f_e^2}{2}(g_{t(e)}A_{t(e)}^\mu-g_{o(e)}A_{o(e)}^\mu+\partial^\mu
a_e)^2-\frac{1}{2}\sum_{e\in
E}(\partial^\mu\rho_e)^2 \notag \\
&~&-\sum_{e,e'\in
E}\sum_{v\in
V}f_e\rho_e(E^T)_{ev}g^2_v(E)_{ve'}f_{e'}
\rho_{e'}+\frac{1}{2}\sum_{v\in
V}\left\{D_v -\frac{\sqrt{2}}{2}g_v\sum_{e\in
E}
(E)_{ve}f_e\rho_e\right\}^2.
\end{eqnarray}

Now one can easily find the mass matrices for vectors and scalars:
\begin{equation}
M^2_A=2GEF^2E^TG=2(GEF)(GEF)^T\,,\quad
M^2_\rho=2FE^TG^2EF=2(GEF)^T(GEF)\,,
\end{equation}
where $E$ is defined as (\ref{Ei}) while $G$ and $F$ are given by
(\ref{GF}). Massless scalar fields are absent if and only if the graph is
a tree graph.
The mass spectrum of the scalar
fields is the same as the one for the vector fields except for zero
modes.\footnote{It is well known that two square matrices $AB$ and $BA$
have the same eigenvalues up to zero modes. See Appendix~\ref{ABBA}.}

The fermionic part of the theory is found to be
\begin{eqnarray}
{\cal L}_f&=&-i\sum_{v\in
V}\frac{1}{g_v^2}\lambda_v\sigma^\mu\partial_\mu\bar\lambda_v
-i\sum_{e\in E}2f_e^2
\chi_e\sigma^\mu\partial_\mu\bar\chi_e  \nonumber \\
&~&+\sum_{e\in
E}2f_e^2\left[\chi_e(\lambda_{t(e)}-\lambda_{o(e)})+h.c.\right]\,,
\end{eqnarray}
and can be rescaled as
\begin{eqnarray}
{\cal L}_f&=&-i\sum_{v\in
V}\lambda_v\sigma^\mu\partial_\mu\bar\lambda_v
-i\sum_{e\in E}
\chi_e\sigma^\mu\partial_\mu\bar\chi_e \nonumber \\
&~&-\sum_{e\in
E}\sum_{v\in
V}\sqrt{2}\left[f_e\chi_e(E^T)_{ev}g_v\lambda_v+h.c.\right].
\end{eqnarray}
Here $\lambda_v$ and $\chi_e$ are Weyl spinor fields contained in $V_v$
and $S_e$, respectively.

One will find the mass matrices for fermions after rescaling the fields:
\begin{equation}
M^2_\lambda=2GEF^2E^TG=2(GEF)(GEF)^T\,,\quad
M^2_\chi=2FE^TG^2EF=2(GEF)^T(GEF).
\end{equation}
Note that the fermions $\lambda$ and $\chi$ form Dirac fields for
massive modes. Also note that all field contents are neutral as well as
free from interactions.

%\newpage
\section{multi-vector, multi-Higgs model}
\subsection{general construction}

We will construct the model that the symmetry $[U(1)]^p$ is {\it
spontaneously} broken to $U(1)$. Therefore we will not use the
Stueckelberg fields but the Higgs fields. 

As the model in the previous section, we consider vector
superfields on vertices and suppose that
$U(1)$ transformation is defined at each vertex.
Moreover in the present case, we introduce a `bi-charged' scalar field
$\Sigma$ on each edge, which is transformed under two $U(1)$ symmetries
as~\footnote{Note that the transformation law for
$\Sigma_e$ is the same as that for $e^{2S_e}$ in the previous section.},
\begin{equation}
\Sigma_e\rightarrow
e^{-2i\Lambda_{t(e)}}\,\Sigma_e\,e^{2i\Lambda_{o(e)}}\, .
%\quad\left(\Sigma_e^\dagger\rightarrow
%e^{-2i\Lambda_{o(e)}^\dagger}\,\Sigma_e^\dagger\,e^{2i\Lambda_{t(e)}^\dagger}\,
%,\right)
\end{equation}
Now we get the $[U(1)]^{p}$ invariant supersymmetric multi-vector,
multi-`Higgs' model on a graph governed by the following Lagrangian:
\begin{eqnarray}
{\cal L}&=&\frac{1}{4}\sum_{v\in
V}\left(\left.W^\alpha_vW^v_{\alpha}\right|_{\theta\theta}+\left.
\overline{W}^v_{\dot{\alpha}}
\overline{W}^{\dot{\alpha}}_v\right|_{\bar\theta\bar\theta}\right)+\sum_{e\in
E}\,\left.\overline\Sigma_e\,e^{2gV_{t(e)}}\Sigma_e\,e^{-2gV_{o(e)}}
\right|_{\theta\theta\bar\theta\bar\theta} \nonumber \\
&-&2g\sum_{e\in
V}\,\left.\zeta_e(V_{t(e)}-V_{o(e)})
\right|_{\theta\theta\bar\theta\bar\theta}\, ,
\end{eqnarray}
where we rescale the gauge coupling constant to be seen explicitly.
The Fayet-Illiopoulos terms are chosen so that they are similar to those
in the model of the previous section, when $\zeta_e\approx f_e^2$.%
\footnote{In most general cases, we can choose the Fayet-Illiopoulos
(FI) terms as $\sim\sum_v \zeta_v V_v$. We would like to study aspects of
(gauge and/or super-) symmetry breakdown with the general FI terms
elsewhere.}  This paper will not go into the issue about anomaly and deal
with only classical aspects of the model.
%\footnote{Then, we simply need oppositely
%gauge coupled superfields
%$\tilde\Sigma_e$, or need a few other (sterile?) fields because a gauge
%field is connected to several fermions with both sign of charge.}???

The bosonic part of the Lagrangian reads
\begin{eqnarray}
{\cal L}_b&=&-\frac{1}{4}\sum_{v\in
V}F_{\mu\nu}^vF^{\mu\nu}_v+\frac{1}{2}\sum_{v\in
V}D^2_v-\sum_{e\in
E}({\cal D}_\mu\sigma_e)^\dagger({\cal D}^\mu\sigma_e) \nonumber \\
&+&\sum_{e\in
E}F^\dagger_{\Sigma\,e}F_{\Sigma\,e}
+g\sum_{e\in
E}(D_{t(e)}-D_{o(e)})\sigma^\dagger_e\sigma_e
-g\sum_{e\in
V}(D_{t(e)}-D_{o(e)})\zeta_e\, ,
\end{eqnarray}
where the covariant derivative is
\begin{equation}
{\cal
D}^{\mu}\sigma_e=(\partial^{\mu}+igA^{\mu}_{t(e)}-igA^{\mu}_{o(e)})
\sigma_e\,.
\end{equation}

By use of the {incidence matrix} of the graph, we rewrite the above
Lagrangian as
\begin{eqnarray}
{\cal L}_b&=&-\frac{1}{4}\sum_{v\in
V}F_{\mu\nu}^vF^{\mu\nu}_v+\frac{1}{2}\sum_{v\in
V}D^2_v-\sum_{e\in
E}({\cal D}_\mu\sigma_e)^\dagger({\cal D}^\mu\sigma_e) \nonumber \\
&+&\sum_{e\in
E}F^\dagger_{\Sigma\,e}F_{\Sigma\,e}
-g\sum_{e\in
E}(\sigma^\dagger_e\sigma_e-\zeta_e)(E^TD)_e\,.
\end{eqnarray}
Substituting the equation of motion for the auxiliary fields
\begin{equation}
F_{\Sigma\,e}=0\quad {\rm and}\quad D_v=g\sum_{e\in
E}(\sigma^\dagger_e\sigma_e-\zeta_e)(E^T)_v\,,
\end{equation}
into the bosonic Lagrangian, we obtain
\begin{eqnarray}
{\cal L}_b&=&-\frac{1}{4}\sum_{v\in
V}F_{\mu\nu}^vF^{\mu\nu}_v-\sum_{e\in
E}({\cal D}_\mu\sigma_e)^\dagger({\cal D}^\mu\sigma_e) \nonumber \\
&-&\frac{g^2}{2}\sum_{e,e'\in
E}(\sigma^\dagger_e\sigma_e-\zeta_e)(E^TE)_{ee'}
(\sigma^\dagger_{e'}\sigma_{e'}-\zeta_{e'})\,.
\label{LbSSB}
\end{eqnarray}
Note that $E^TE$ is a $(q, q)$ matrix.

%\newpage
\subsection{example: $P_3$}

\begin{figure}[h]
\begin{center}
\includegraphics[height=2.5cm]
{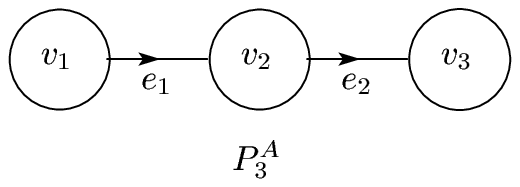}
\vspace{5mm}
\includegraphics[height=2.5cm]
{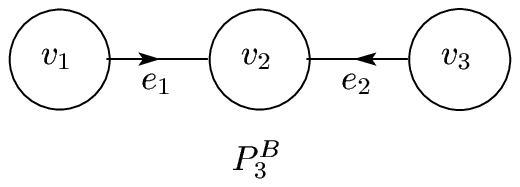}
\end{center}
\caption{$P_3$: the path graph with three vertices.
There are two substantially different graphs. 
They have the different incidence matrices.}
\label{pathgraph002+}
\end{figure}

The structure of the model depends on the incidence matrix of the graph.
For a simple example, let us consider the path graph with three vertices,
$P_3$.

The incidence matrix depends on the orientation of edges. For instance,
two cases can be considered as follows:\footnote{Obviously the overall
sign of the incidence matrix is irrelevant.}
\begin{equation}
({E_A})_{ve}=\left(
\begin{array}{cc}
1 & 0\\
-1 & 1\\
0 & -1
\end{array}
\right)\,,\quad
({E_B})_{ve}=\left(
\begin{array}{cc}
1 & 0\\
-1 & -1\\
0 & 1
\end{array}
\right)\,,
\end{equation}
where $E_A$ is the incidence matrix of $P_{3}^A$ and $E_B$ is
the one of $P_{3}^B$.

Interestingly, the following matrix is independent of the edge
orientation:
\begin{equation}
{E_AE_A^T}={E_BE_B^T}=\left(
\begin{array}{ccc}
1 & -1 & 0\\
-1 & 2 & -1\\
0 & -1 & 1
\end{array}
\right)\equiv \Delta\,.
\end{equation}
This is known as the {graph Laplacian}.

On the other hand, we find
\begin{equation}
{E_A^TE_A}=\left(
\begin{array}{cc}
~2 & -1 \\
-1 & ~2 
\end{array}
\right)\,,\quad
{E_B^TE_B}=\left(
\begin{array}{cc}
~2 & ~1 \\
~1 & ~2 
\end{array}
\right)\,.
\end{equation}
Therefore the shape of the `Higgs' potential in Eq.~(\ref{LbSSB}) depends
on the edge orientation.

%%%%%%%%%%%%%%%%%%%%%%%%%%%
%\begin{wrapfigure}{r}{5cm}
\begin{figure}[h]
\centering
\includegraphics[height=5cm]
{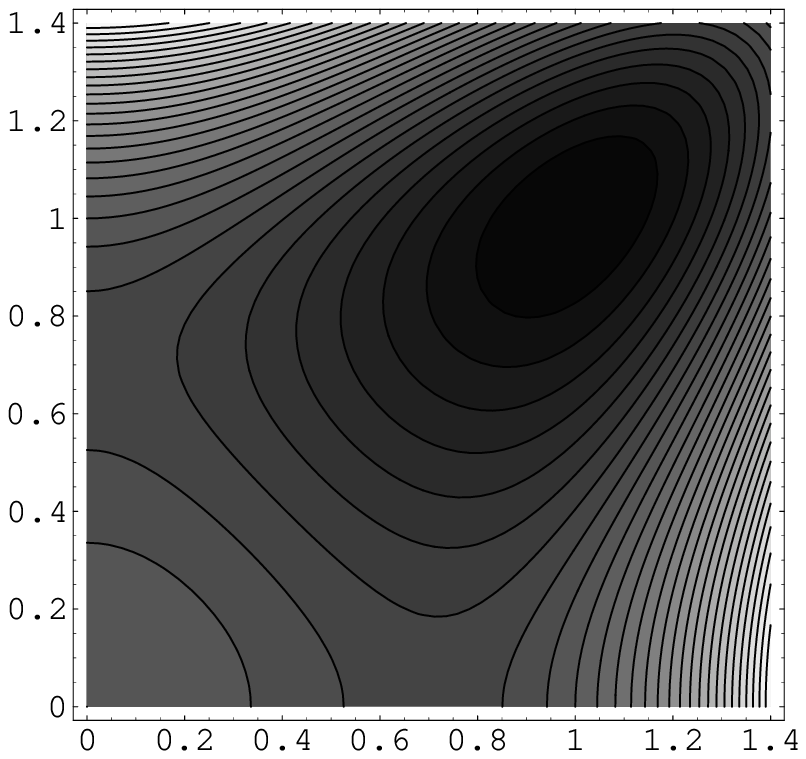}
\qquad\qquad
\includegraphics[height=5cm]
{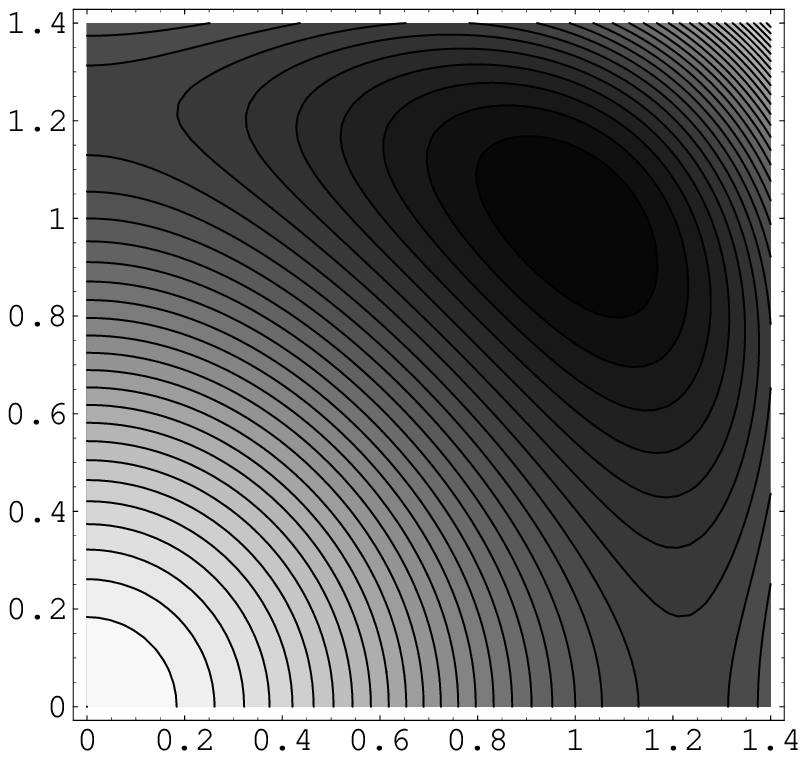}
\caption{%
Contour plots of scalar potentials for the models based on $P^A_3$
(left) and on
$P^B_3$ (right), respectively. In both plots, potentials are normalized by 
$g^2f^4$, the contour spacing is $0.1$, and the horizontal axis
indicates
$|\sigma_1|/f$ while the vertical axis indicates $|\sigma_2|/f$.}
\label{figpot}
\end{figure}
%\end{wrapfigure}
%%%%%%%%%%%%%%%%%%%%%%%%%%%

Figure~\ref{figpot} illustrates the contour plots of the potentials in
Eq.~(\ref{LbSSB}) for the graphs $P^A_3$ and $P^B_3$.

\subsection{mass matrices for bosonic and fermionic fields}
Individually different gauge coupling constants will also be considered.
The consequence of such consideration 
forces the bosonic part of the Lagrangian to be
\begin{eqnarray}
{\cal L}_b&=&-\frac{1}{4}\sum_{v\in
V}F_{\mu\nu}^vF^{\mu\nu}_v-\sum_{e\in
E}({\cal D}_\mu\sigma_e)^\dagger({\cal D}^\mu\sigma_e)\nonumber \\ %\notag
&-&\frac{1}{2}\sum_{e,e'\in
E}\sum_{v\in
V}(\sigma^\dagger_e\sigma_e-\zeta_e)(E^T)_{ev}g^2_v(E)_{ve'}
(\sigma^\dagger_{e'}\sigma_{e'}-\zeta_{e'})\,,
\end{eqnarray}
with
\begin{equation}
{\cal
D}^{\mu}\sigma_e=(\partial^{\mu}+ig_{t(e)}A^{\mu}_{t(e)}
-ig_{o(e)}A^{\mu}_{o(e)})\sigma_e\,.
\end{equation}
Here we assume that all $\zeta_e$ are positive and $\sqrt{\zeta_e}=f_e$. 
Thus the vacuum expectation value for $|\sigma_e|$ is $f_e$ and physical
scalar fields should be considered as the linear combinations of 
$|\sigma_e|-f_e$. Each phase part of a to-be massive scalar field is
eaten by a vector field through the Higgs mechanism. Then the
$(mass)^2$ matrices
$M_V^2$ for vector fields and $M_S^2$ for scalar fields in this case are
\begin{equation}
M^2_V=2GEF^2E^TG=2(GEF)(GEF)^T\,,\quad
M^2_S=2FE^TG^2EF=2(GEF)^T(GEF)\,,
\end{equation}
where the matrices that appeared in the above equations are the same as
(\ref{Ei}) and (\ref{GF}).

Although the shape of the potential with respect to $|\sigma_e|$ depends
on the orientation of edges in the graph, the mass spectrum of the scalar
fields is the same as the one for the vector fields except for zero
modes, similarly to the model in the previous section.

The number of the moduli of the potential is $q-p+1$ for a general graph.
This is equal to the number of independent closed circuits in the
graph.\footnote{If
$q-p+1>0$, the graph has a closed circuit $C(G)$. It is possible that we
add the term like
$\sum_{\{e_1,e_2,e_3\}\in C(G)}\Sigma_{e_1}\Sigma_{e_2}\Sigma_{e_3}$
to the Lagrangian to give the scalar masses.} 
For tree graphs, the
vacuum expectation values of $\sigma_e$ are determined rigidly if all
$\zeta_e$ are positive.

The fermionic part of the Lagrangian is
\begin{eqnarray}
{\cal L}_f&=&-i\sum_{v\in
V}\lambda_v\sigma^\mu\partial_\mu\bar\lambda_v
-i\sum_{e\in
E}\psi_e\sigma^\mu{\cal D}_\mu\bar\psi_e \nonumber \\
&+&
i\sqrt{2}\sum_{e\in
E}(\sigma_e\bar\psi_e(E^T)_{ev}g_v\bar\lambda_v-
\sigma_e^\dagger\psi_e(E^T)_{ev}g_v\lambda)\,,
\end{eqnarray}
where 
$\lambda_v$ and $\psi_e$ are Weyl spinor fields contained in $V_v$
and $\Sigma_e$, respectively. The covariant derivative on $\psi_e$ is
defined as
${\cal
D}^{\mu}\psi_e=(\partial^{\mu}+ig_{t(e)}A^{\mu}_{t(e)}
-ig_{o(e)}A^{\mu}_{o(e)})\psi_e$.
Substituting the vacuum expectation values $\langle\sigma_e\rangle=f_e$,
we find
\begin{equation}
M^2_\lambda=2(GEF)(GEF)^T\,,\quad
M^2_\psi=2(GEF)^T(GEF)\,.
\end{equation}
Since SUSY is unbroken, the bosonic and fermionic spectra are 
the same.

In this paper, we have considered models with unbroken SUSY.
The model with `partially' broken SUSY is interesting, for some
$\zeta_e<0$. The present analysis will not go into such models.

\section{vortex solution}
It is well known that the vortex solution can be found in the
Abelian-Higgs model~\cite{AHM}. In many papers, the solution is used as
a simple model for a cosmic string~\cite{CosmicString}. We consider the
vortex-type solutions in our model described in the previous section.

Although an academic interest in our toy model is an important motivation
for the following study, we also think that topological configurations are
a key ingredient in recent studies in theoretical physics. A possibility
is expected that a similar model provides an example of a complicated
brane/string system. In the present paper, anyway, we study only simple
vortex in our theory and their generalizations and possible applications
to particle physics and cosmology are left for future work.

Moreover we will consider only tree graphs as the bases of models.

\subsection{Bogomolnyi equation}
In the Abelian-Higgs model, the vortex solution is well known~\cite{AHM}.
Moreover, it is known~\cite{Edelstein} that supersymmetric $U(1)$ theory satisfies the
Bogomolnyi condition~\cite{sB}. Because our model is also supersymmetric,
the Bogomolnyi condition can be found. The equations of motion can be
reduced to the following two sets of equations:
\begin{equation}
F^{ij}_v=\mp
\varepsilon^{ij}g_v\sum_{e\in E}(E)_{ve}(|\sigma_e|^2-\zeta_e)\,,
\label{bogo1}
\end{equation}
and
\begin{equation}
{\cal D}^i\sigma_e=\mp i \varepsilon^{ij}{\cal D}_j\sigma_e\,,
\label{bogo2}
\end{equation}
where $i, j$ denote two spatial directions and $\varepsilon^{ij}$ is the
antisymmetric tensor.

These equations are the Bogomolnyi equations.

The energy per unit length of a vortex string can be written as
\begin{eqnarray}
{\cal E}&=&\int d^2x\left[\frac{1}{4}\sum_{v\in
V}F_{ij}^vF^{ij}_v+\sum_{e\in
E}({\cal D}_i\sigma_e)^\dagger({\cal D}^i\sigma_e)\right. \nonumber \\
&~&\qquad\qquad+\left.\frac{1}{2}\sum_{e,e'\in
E}\sum_{v\in
V}(|\sigma_e|^2-\zeta_e)(E^T)_{ev}g^2_v(E)_{ve'}
(|\sigma_{e'}|^2-\zeta_{e'})\right] \\
&=&\int d^2x \left[\frac{1}{4}\sum_{v\in V}\left\{F^{ij}_v\pm
\varepsilon^{ij}g_v(E)_{ve}(|\sigma_e|^2-\zeta_e)\right\}^2+
\frac{1}{2}\sum_{e\in
E}
\left|{\cal D}^i\sigma_e\pm
i\varepsilon^{ij}{\cal D}_j\sigma_e\right|^2
\right. \nonumber \\
&~&\qquad\qquad\pm\left.\left\{\sum_{v\in
V}\sum_{e\in E}\frac{1}{2}\varepsilon_{ij}F^{ij}_v
g_v(E)_{ve}\zeta_e-i\sum_{e\in
E}\varepsilon^{ij}\partial_i(\sigma_e^\dagger{\cal
D}_j\sigma_e)\right\}\right].
\end{eqnarray}
For a solution of finite energy density,  ${\cal D}_i\sigma_e$ is equal to zero at spatial infinity. 
If the asymptotic behavior of $\sigma_e$ is expressed by the
azimuthal angle $\varphi$ and an integer $n_e$, {\it i.e.}
$\sigma_e\rightarrow\sqrt{\zeta_e}e^{in_e\varphi}$,
the condition tells $(E^Tg_vA_i^v)_e\rightarrow n_e\partial_i\varphi$,
and then  $\int d^2x\, (E^T\varepsilon_{ij}g_vF^{ij}_v)_e=4\pi n_e$.
Therefore the energy density becomes
\begin{eqnarray}
{\cal E}&=&\int d^2x \left[\frac{1}{4}\sum_{v\in V}\left\{F^{ij}_v\pm
\varepsilon^{ij}g_v(E)_{ve}(|\sigma_e|^2-\zeta_e)\right\}^2+
\frac{1}{2}\sum_{e\in
E}
\left|{\cal D}^i\sigma_e\pm
i\varepsilon^{ij}{\cal D}_j\sigma_e\right|^2
\right] \nonumber \\ %\notag
&~&\pm 2\pi\sum_{e\in E}|n_e|\zeta_e\,.
\end{eqnarray}

We deal with the lowest bound for the energy density read from this
result. 
%Of course, if all winding number $n_e$ has the same sign,
The vortex solution satisfying the Bogomolnyi equation
(\ref{bogo1},\ref{bogo2}) has the energy density
$2\pi\sum_{e\in E}|n_e|\zeta_e$.%
\footnote{Because of the presence of many fields, non-Bogomolnyi 
configuration may have lower energy ({\it i.e.}, the Bogomolnyi solution
may correspond to a local minimum).}

\subsection{Bogomolnyi vortices and SUSY}
It is well known that the SUSY is partially broken in the topological 
background fields. Here we briefly describe the pattern of SUSY breaking
in our model. Notation is indebted in~\cite{SUSY}. According to SUSY, the
variations of the gauginos
$\lambda_e$ are
\begin{equation}
\delta_\epsilon\lambda_v=i\epsilon
D_v+\sigma^{\mu\nu}F_{v\,\mu\nu}\epsilon\,.
\end{equation}
Using the Bogomolnyi equations~(\ref{bogo1}), and assuming the vortex
string lies in the third direction for simplicity, the above variations
are rewritten as
\begin{equation}
\delta_\epsilon\lambda_v=\mp i
F^{12}_v(1\pm\sigma^3)\epsilon\,.
\end{equation}
This means that the half of the SUSY at the vertex is broken in the
presence of the central magnetic flux of the vortex.

The variations of partners of $\sigma_e$ are
\begin{equation}
\delta_\epsilon\psi_e=i\sqrt{2}\bar\epsilon\,\sigma^\mu
{\cal D}_\mu\sigma_e\,,
\end{equation}
where ${\cal
D}^\mu\psi_e\equiv\partial^\mu\psi_e+i((gA)_{t(e)}^\mu-
(gA)_{o(e)}^\mu)\psi_e$. If the vortex string lies in the third direction,
this reduces when the Bogomolnyi equations~(\ref{bogo2}) hold,
\begin{equation}
\delta_\epsilon\psi_e=i\sqrt{2}\bar\epsilon\,\left[\sigma^1
{\cal D}_1\psi_e+\sigma^2
{\cal D}_2\psi_e\right]
=i\sqrt{2}\bar\epsilon\,(\sigma^1\pm i\sigma^2)
{\cal D}_1\psi_e\,.
\end{equation}
We find again that the half of the SUSY at the edge is broken in
the presence of the magnetic flux.

%asymptotic solutions\\

%fermion zero modes\\

%multi-vortex solution?

%\newpage
\subsection{construction of vortices: ansatz}
Next we examine how we can obtain the explicit solutions in
our model. For simplicity, we consider a common gauge coupling constant
$g$ and a single constant $f=\sqrt{\zeta}$. In other words, we consider
the case that $G=g I$ and $F=f I$ (where $I$ is the
identity matrix). Although we cannot tell about most general solutions,
we take ansatz for simple, physically admissible type of vortex
solutions.\footnote{For a reference, we write down the construction of
normal vortex solutions in Appendix~\ref{NV}.} We impose the axially
symmetric ansatz 
\begin{align}
& \sigma_{e} = \rho_e (r)\, e^{i n_e \varphi}\,, \\
& A^v_{\varphi} = P_v (r)\,,
\end{align}
on Bogomolnyi equations.
Here we express the radial coordinate as $r$ and the azimuthal angle as
$\varphi$. The integers $n_e$ are winding numbers. The detailed
calculation is shown in the Appendix~\ref{detail}. We get the following
Bogomolnyi equations,
\begin{align}
& \dfrac{\rho_e'}{\rho_e} = - \dfrac{\left( g \left( E^T P \right) -n
\right)_e}{r}\,, \\ & \dfrac{P_v'}{r} = -g \sum_{e \in E}\left(E
\right)_{ve} \left( \rho^2 - f^2 \right)_e\,,
\end{align}
where the prime ($'$) denotes the derivative with respect to $r$.
These equations are the special case of the Bogomolnyi equations.

%\newpage
\subsection{examples of vortex solutions}
We show some concrete examples for the vortex solution in our model.
To have the vortex solution we restrict the graph structure, or
equivalently, the incident matrix $E$.
Here we also consider configurations with the least winding numbers for
simplicity and for feasibility in physical systems.

We consider here the cases with the single-centered exact solution similar
to the normal vortex. The asymptotic behavior of general cases can be
obtained and is shown in Appendix~\ref{asy}.

\subsubsection{Example 1: $P_2$}
\begin{figure}[h]
\begin{center}
\includegraphics[height=2.5cm]
{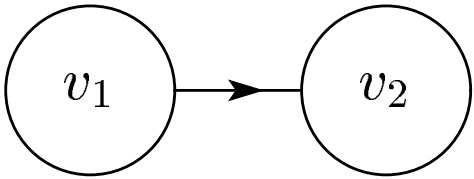}
\end{center}
\caption{$P_2$: the path graph with two vertices.}
\label{pathgraph001}
\end{figure}

The simplest case has two vertices and an edge.
This graph is $P_2$ graph.
We show the graph in FIG.~\ref{pathgraph001}.

In this case, the incidence matrix and its transposed matrix are
\begin{equation}
(E)_{ve}=\begin{pmatrix} 1 \\ -1 \\  \end{pmatrix}, 
\ \ \ \ \left(E^{T}\right)_{ev}=\begin{pmatrix} 1 & -1 \\ 
\end{pmatrix}\,.
\end{equation}
Then considering the Bogomolnyi equations
\begin{align}
\dfrac{P_v'}{r} &= -g \sum_{e \in E} (E)_{ve} \left( \rho^2
- f^2 \right)_e\,, \\
\dfrac{\rho_e'}{\rho_e} &= - \dfrac{\left(g\, E^T P - n \right)_e}{r}\,,
\end{align}
the first one becomes
\begin{align}
\dfrac{P_1'}{r} &= -g \left( \rho^2 - f^2 \right)\,, \\
\dfrac{P_2'}{r} &= +g \left( \rho^2 - f^2 \right)\,. 
\end{align}
Therefore it is necessary to find a set of unique equations that we
suppose the relation $P_1(r)=-P_2(r)$. On the other hand, in the second
equation we notice
\begin{equation}
\sum_{v}(E^T)_{ev} P_v = \begin{pmatrix} 1 & -1 \\
\end{pmatrix}
\begin{pmatrix} 1 \\ -1 \\ \end{pmatrix} P_1 = 2P_1\,.
\end{equation}
So, we get the following equations
\begin{align}
\dfrac{P_1'}{r} &= -g \left( \rho^2 - f^2 \right)\,, \\
\dfrac{\rho'}{\rho} &= - \dfrac{2g P_1 - n}{r}\,.
\end{align}
These equations can be reduced to
\begin{align}
\dfrac{\Tilde{P}'}{x} &= - \left( \Tilde{\rho}^2 - 1 \right)\,, 
\label{AA}\\
\dfrac{\Tilde{\rho}'}{\Tilde{\rho}} &= -
\dfrac{\Tilde{P}-n}{x}\,,
\label{BB}
\end{align}   
if we rescale the variables so that $\Tilde{P}(x)=2gP_1(r)$,
$\Tilde{\rho}(x)={\rho}(r)/{f}$,
$x=\sqrt{2} gf r$, $n=1$  and the prime ($'$) is the derivative with
respect to 
$x$. These equations are precisely same as the normal Bogomolnyi
equations. The normal Bogomolnyi equations is referred in
Appendix~\ref{NV}.

%In the $P_2$ graph, we have the vortex solution when it is satisfied 
%that $P_1(r)=-P_2(r)$. This is the condition of the vortex solution.
The energy per unit length of the straight string is given by $2\pi f^2$
in this case. Generalization to the case with the winding number $n>1$ is
trivial.

\subsubsection{Example 2: $P_3$}
\begin{figure}[h]
\begin{center}
\includegraphics[height=2.5cm]
{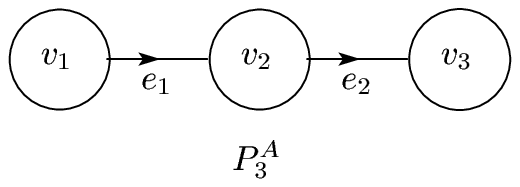}
\vspace{5mm}
\includegraphics[height=2.5cm]
{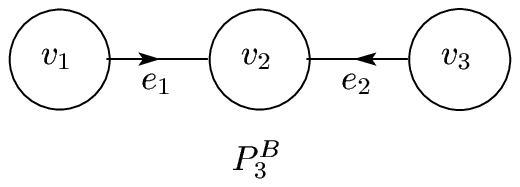}
\end{center}
\caption{The graph $P_{3}^A$ has edges of the same direction while
$P_{3}^B$ has the edges of the different direction.}
\label{pathgraph002}
\end{figure}

We consider the $P_3$ graph, the three-vertex path graph.
In this graph, we consider two patterns of the direction of the edges.
We show these in FIG.~\ref{pathgraph002}.
 
The condition to reduce the Bogomolnyi equations in these cases to the
normal ones (\ref{AA},\ref{BB}) with $\rho_1=\rho_2$ and $n_1=n_2=1$ are 
$P_1(r)=-P_3(r)$ and
$P_2(r)\equiv 0$  in the case with
$P_{3}^A$ while
$P_1(r)=P_3(r)$ and $P_2(r)=-2P_1(r)$  in the case with
$P_{3}^B$.
The necessary scaling is that $\Tilde{P}(x)=gP_1(r)$ and $x=gfr$ in the
case with
$P_3^A$ while $\Tilde{P}(x)=3gP_1(r)$ and $x=\sqrt{3}gfr$ in the case with
$P_3^B$. The energy density takes the same value $2\pi f^2 (1+1)=4\pi
f^2$ in both cases.

\subsubsection{Example 3: $K_{1,N}$}
\begin{figure}[h]
\begin{center}
\includegraphics[height=5cm]
{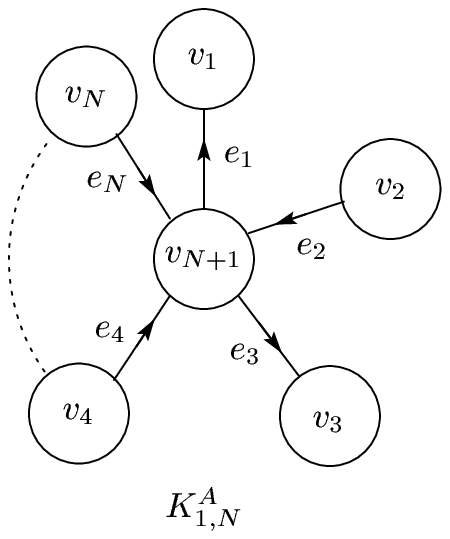}
\includegraphics[height=5cm]
{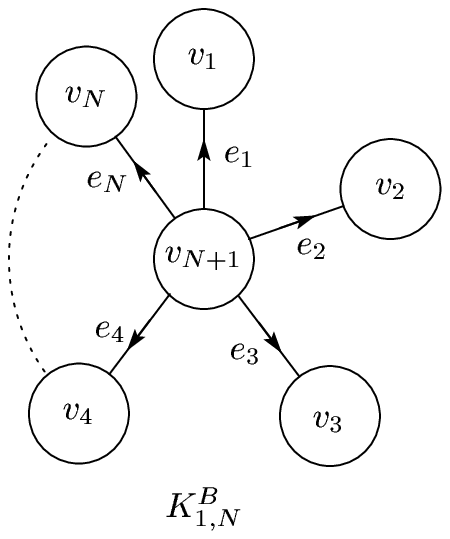}
\end{center}
\caption{The star graphs, $K_{1,N}^A$ and $K_{1,N}^B$.}
\label{stargraph00N}
\end{figure}
We consider another tree graph, the star graph $K_{1,N}$.
In the star graph, $v_{N+1}$ is adjacent to all the other vertices and 
no extra edge exists.
We recognize two types of edges.
One is the edge whose origin is $v_{N+1}$, another edge is one whose
 terminus is $v_{N+1}$. We call the edge of the first type is
$e_{o}$, the one of the second type is $e_{t}$. 

We heuristically find the cases that we get the vortex
solution similar to the normal one with
$\rho_1=\rho_2=\cdots=\rho_N=\rho_{N+1}$:  Here two cases are shown where
the number of edges belonging to two types are
\begin{align}
K_{1,N}^A:\quad&\# e_{o}= \# e_{t}=N/2\,, \label{same} \\
K_{1,N}^B:\quad&\# e_{o} = N\quad{\rm and}\quad \ \#
e_{t}=0\,,~or~vice~versa\,,
\label{diff} 
\end{align}
where, off course, $N$ is considered to be even in the case $A$.
The graphs of two types are shown in FIG.~\ref{stargraph00N}.

The incidence matrix of $K^A_{1,N}$ (where $N$ is even) is $(N+1, N)$
matrix given by
\begin{equation}
(E_A)_{ve}=\left(
\begin{array}{ccccccc}
-1 &  0 &  0&\cdots&0\\
 0 &  1 &  0&\cdots&0\\
 0 &  0 & -1&\cdots&0\\
\vdots &\vdots &\vdots &\ddots&\vdots\\
 0 &  0 &  0&\cdots& 1\\
 1 & -1 &  1&\cdots& -1
\end{array}
\right)\,,
\end{equation}
while the incidence matrix of $K^B_{1,N}$ is
\begin{equation}
(E_B)_{ve}=\left(
\begin{array}{ccccccc}
-1 & 0 & 0&\cdots&0\\
0 & -1 & 0&\cdots&0\\
0 & 0 & -1&\cdots&0\\
\vdots &\vdots &\vdots &\ddots&\vdots\\
0 & 0 & 0&\cdots&-1\\
1 & 1 & 1&\cdots& 1
\end{array}
\right)\,.
\end{equation}

We found these patterns by extending the analysis of getting the vortex
solution in the case with
$P_3$ graph shown previously, because $K_{1,2}$ is the same as $P_{3}$.

In the first case (\ref{same}), we have vortex solutions if
$P_{2\ell-1}(r)=-P_{2m}(r)$ ($\ell,m$ are positive integers and $\ell, m
\le \frac{N}{2} $) and $P_{N+1}\equiv 0$.   In the second case
(\ref{diff}), we have the solutions if
$P_1(r)=P_2(r)= \cdots = P_N(r)$ and $P_{N+1}(r)=-NP_1(r)$.  
In both cases the energy density is found to be $2\pi N f^2$ if all the
winding numbers are unity.

\subsubsection{Inclusion of `no winding scalar edge'}
\begin{figure}[h]
\begin{center}
\includegraphics[width=5cm]
{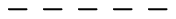}
\end{center}
\caption{This dashed line means  that  $\rho_e
\equiv f$ on this edge, `no winding scalar edge'.}
\label{line02}
\end{figure}

In the previous two examples, all `Higgs' scalars have nonzero winding
number. Conversely we consider that there is an edge where the
assigned scalar has no winding number, thus
$\rho_e \equiv f$ at the edge. 
We use
the dashed line to express such an edge, as in FIG.~\ref{line02}. 

For a constant $\rho_e$, $P_{o(e)}(r)\equiv P_{t(e)}(r)$ holds
everywhere.%
\footnote{Thus the orientation of the edge is irrelevant (so, there is no
arrow assigned to the dashed line).} 
%Including the condition $(\rho^2 - f^2)_{e}=0$,
%the vortex solution is  the combination of the graph having a vortex
%solution. 
Suppose that one have already constructed the vortex solution in a
certain model with specific graph structure. The one might duplicate
the solution  and the graph. One may connect the identical vertices of
the original and copy of the graph by `no winding scalar edge'.
The number of such connection is arbitrary.
This method can be applied to the case with two different models and
solutions, if one finds the same functional form of $P_v(r)$ in each
model.
Of course more than two vertices can be connected if $P_v$ is common at
all vertices.

\subsubsection{Example 4: $P_{4}$}
\begin{figure}[h]%o___o...o___o
\begin{center}
\includegraphics[width=10cm]
{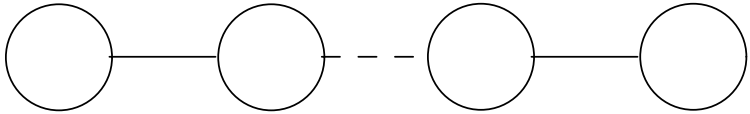}
\end{center}
\caption{$P_4$ graph consists of two $P_2$ and an edge.}
\label{pathgraph001+1}
\end{figure}
\begin{figure}[h]%o_<_o...o_>_o
\begin{center}
\includegraphics[width=10cm]
{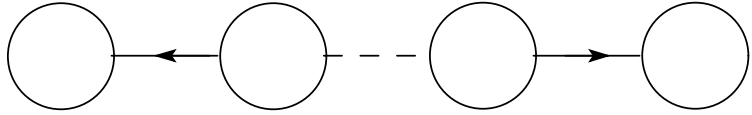}
\end{center}
\caption{$P_4$ graph whose edge direction is
left-right symmetric with respect to the dashed edge. Each of edge
directions is outgoing with respect to the dashed edge.}
\label{pathgraph001+1-1}
\end{figure}
\begin{figure}[h]%o_>_o...o_<_o
\begin{center}
\includegraphics[width=10cm]
{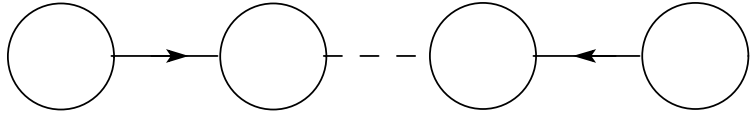}
\end{center}
\caption{$P_4$ graph. Each of edge direction
is incoming with respect to the dashed edge.}
\label{pathgraph001+1-2}
\end{figure}

We consider the $P_4$ graph. The graph $P_4$ has two $P_2$ as subgraphs
and is shown in FIG.~\ref{pathgraph001+1}. 
We do not show the direction of the edge in this graph. 
This graph has a left-right symmetry with respect to the dashed edge.
This symmetry is connected with the winding number of each vector fields. 
The vector fields at the both ends of the dashed line must be described by
an identical function.  
For this reason, we should impose the left-right
symmetry to the direction of edges.  
In the $P_4$ case, we find two types of the edge orientation graph
for admitting the normal vortex solutions, shown in
FIG.~\ref{pathgraph001+1-1} and FIG.~\ref{pathgraph001+1-2}.
In the similar way, we consider the model based on $P_{2\ell}$ with normal
vortex solutions.

\subsubsection{Example 5: $P_{6}$}
\begin{figure}[h]%o___o___o...o___o___o x4
\begin{center}
\includegraphics[height=5cm]
{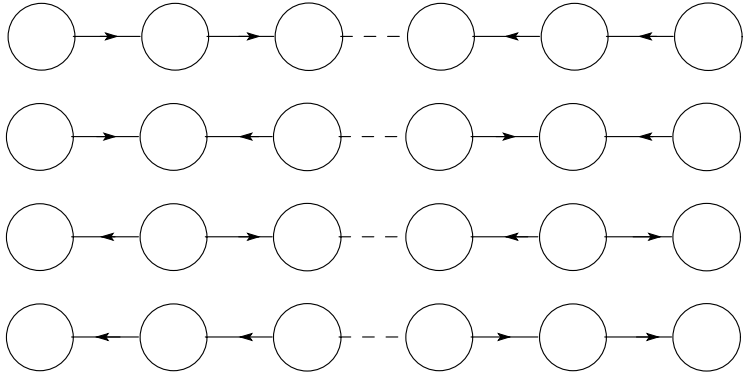}
\end{center}
\caption{There are four types of the $P_6$ graph consisting of two $P_3$.}
\label{pathgraph002+2-1}
\end{figure}

The graph $P_6$ has three $P_2$ as subgraphs. We study the model based on
$P_6$ and their standard solution in the above-mentioned way.
\begin{figure}[h]%o___o___o...o___o___o
\begin{center}
\includegraphics[width=15cm]
{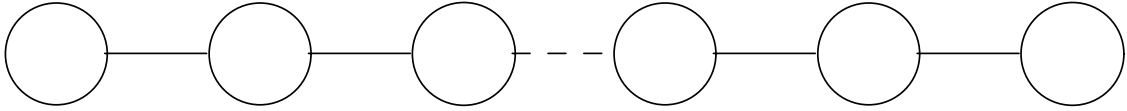}
\end{center}
\caption{$P_6$ graph, which includes two $P_3$ as subgraphs.}
\label{pathgraph002+2}
\end{figure}
In addition, $P_6$ has two $P_3$ as subgraphs.
Similarly to the case with $P_4$, we can consider the $P_6$ graph as two
subgraphs connected by an edge. We exhibit the $P_6$ graph in
FIG.~\ref{pathgraph002+2}. We have the left-right symmetry with respect
to the dashed edge also in this case. We classify four types of the graph
in terms of the direction of the  edges as in
FIG.~\ref{pathgraph002+2-1}. 
In the similar way, we can consider the $P_{3\ell}$ graph, and
associated models and solutions. 

\subsubsection{Example 6}
\begin{figure}[h]%K_{1,N} x2
\begin{center}
\includegraphics[height=5cm]
{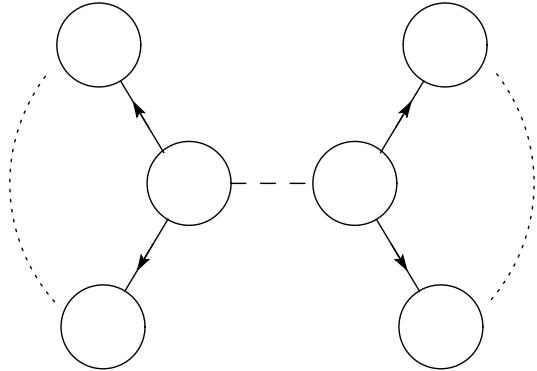}
\end{center}
\caption{The graph consisting of two $K_{1,N}$ connected by the dashed
edge.}
\label{stargraph00N+N}
\end{figure}

We can connect two $K_{1,N}$ graphs by the dashed edge as in
FIG.~\ref{stargraph00N+N}.
As this example, we can find the graph structure admitting the normal
vortex solutions.

%\newpage
\section{Conclusion and Outlook}
We have generalized DD into GDD and introduced SUSY to GDD in the
Abelian theory. A multi-Abelian-Higgs model has been studied as a further
generalization. After getting the Bogomolnyi equations, we
explicitly  constructed vortex solutions of the normal type. To get the
vortex solution, we restricted the graph structure to the special cases
shown in the previous section.  We showed some examples
for the graph which has the normal vortex solution.

We have left the following aspects of the multi-Abelian-Higgs models for
future work. First, we discussed single-centered vortex in the present
paper.  The possibility of multi-vortex solution~\cite{Weinberg1} is an
important subject to study. Next, in this paper, we mainly considered
tree graphs. If we take general graph structures as the bases of
multi-Abelian-Higgs models, we have scalar potentials with (many) flat
direction of the lowest energy.  The appearance of moduli is the feature
of supersymmetric theories and the vortex solution in such a model is
crucial for phenomenological models~\cite{Davis}.  At the same time, the
quantum corrections might become essential. The  generalization of the
method in
\cite{RNW1} will be useful to investigate the quantum effects
about vortices. Finally, because our model contains several fields, the
possibility of different types of topological defects, such as
`rings'~\cite{Doudoulakis}, must be examined.

We considered the Abelian gauge theory in GDD as well as
multi-Higgs models. We are also interested in the non-Abelian theory
because the three-site  Higgsless model is based on the $[SU(2)]^2
\otimes U(1)$ gauge theory. While we considered vortices in the Abelian
gauge theory in this paper, on the other hand there exist monopoles in
the non-Abelian gauge theory. As the future works, we wish to incorporate
monopoles, superfields and GDD into non-Abelian theory as some toy models
for the Higgsless model.

%the normalized Laplacian matrix.\\

%lower dimensional SUSY higher dimensional SUSY\\

\section*{Acknowledgement}
We would like to thank T. Hanada
 for useful comments.

\appendix

%%%%%%%%%%%%%%%
%%%%%%%%%%%%%%%
\section{contents of superfields}\label{cSF}
In this Appendix, we collect the superfields and their component fields.
See the reference~\cite{SUSY}. 
\subsection{vector superfield}
\vspace{-10mm}
\begin{equation}
V_v=-\theta\sigma_\mu\bar\theta
A^\mu_v+i\theta\theta\bar\theta\bar\lambda_v
-i\bar\theta\bar\theta\theta\lambda_v+
\frac{1}{2}\theta\theta\bar\theta\bar\theta D_v\, .
\end{equation}
This satisfies
\begin{equation}
V_v^2=-
\frac{1}{2}\theta\theta\bar\theta\bar\theta A_\mu^vA^\mu_v\, ,\quad
V_v^3=0\,.
\end{equation}

\subsection{chiral superfield (Stueckelberg superfield)}
\vspace{-10mm}
\begin{eqnarray}
S_e&=&\frac{1}{2}(\rho_e+ia_e)+\theta\chi_e+i\theta\sigma^\mu\bar\theta
\frac{1}{2}(\partial_\mu\rho_e+i\partial_\mu
a_e) \nonumber \\
&~&+\theta\theta F_{Se}+\frac{i}{2}
\theta\theta\bar\theta\bar\sigma^\mu\partial_\mu\chi_e+
\frac{1}{8}\theta\theta\bar\theta\bar\theta(\rho_e+i a_e)\, ,
\end{eqnarray}
\begin{eqnarray}
S_e+\overline
S_e&=&\rho_e+\theta\chi_e+\bar\theta\bar\chi_e-\theta\sigma^\mu\bar\theta
\partial_\mu a_e+\theta\theta F_{Se}+\bar\theta\bar\theta
F^\dagger_{Se}\nonumber
\\ &~&+\frac{i}{2}
\theta\theta\bar\theta\bar\sigma^\mu\partial_\mu\chi_e
+\frac{i}{2}
\bar\theta\bar\theta\theta\sigma^\mu\partial_\mu\bar\chi_e+
\frac{1}{4}\theta\theta\bar\theta\bar\theta\rho_e\, .
\end{eqnarray}

\subsection{chiral superfield (Higgs superfield)}
\vspace{-10mm}
\begin{eqnarray}
\Sigma_e&=&\sigma_e+\sqrt{2}\theta\psi_e+i\theta\sigma^\mu\bar\theta
\partial_\mu\sigma_e \nonumber \\
&~&+\theta\theta F_{\Sigma e}+\frac{i}{\sqrt{2}}
\theta\theta\bar\theta\bar\sigma^\mu\partial_\mu\psi_e+
\frac{1}{4}\theta\theta\bar\theta\bar\theta(\sigma_e)\, .
\end{eqnarray}

%B
\section{The eigenvalues of matrices $AB$ and $BA$}\label{ABBA}
Let $A$ be a $(p,q)$ matrix and $B$ be a $(q,p)$ matrix.
Then $(p+q, p+q)$ matrices $U$ and $V$ are defined as
\begin{equation}
U=\left(
\begin{array}{cc}
I_p & A\\
B & xI_q
\end{array}
\right)\,,\quad
V=\left(
\begin{array}{cc}
xI_p & -A\\
0_{qp} & I_q
\end{array}
\right)\,,
\end{equation}
where $I_p$ is the $(p,p)$ identity matrix while $0_{qp}$ is the
$(q,p)$ matrix all of which elements are zero.

The products of two matrices are
\begin{equation}
UV=\left(
\begin{array}{cc}
xI_p & 0_{pq}\\
xB & xI_q-BA
\end{array}
\right)\,,\quad
VU=\left(
\begin{array}{cc}
xI_p-AB & 0_{pq}\\
B & xI_q
\end{array}
\right)\,.
\end{equation}
Because $\det UV=\det VU$, the eigenvalues of  $AB$ and $BA$ are equal,
except for zero eigenvalues.

%\newpage
\section{The normal vortex in Abelian-Higgs model}\label{NV}
The Ginzburg-Landau theory is used as a macroscopic theory of the 
superconductivity. That is nonrelativistic theory, and we know an
Abelian-Higgs model as the relativistic version of the Ginzburg-Landau
theory. This model includes the normal vortex solution.  In this paper we
distinguish  the vortex solution of the Abelian-Higgs model from
the vortex solutions of our multi-Abelian-Higgs models, by using the word
``normal''.
   
In the Abelian-Higgs model, the Lagrangian density is
\begin{equation}
\mathcal{L}= - \dfrac{1}{4} F^{\mu\nu}F_{\mu\nu}- \left| D_{\mu} \sigma
\right|^2 -\dfrac{1}{2}g^2 \left( \sigma^2 - f^2 \right)^2\,,
\end{equation}
where $F_{\mu\nu}=\partial_\mu A_\nu-\partial_\nu A_\mu$ is a field
strength of the Abelian gauge field
$A_{\mu}$,
$\sigma$ is a complex scalar field and $f$ is its vacuum expectation 
value $\langle\sigma \rangle=f$.
$D_{\mu}\sigma$ is the covariant derivative of the scalar field
\begin{equation}
D_{\mu} \sigma = \partial_{\mu} \sigma + i g A_{\mu} \sigma,
\end{equation}
where $g$ is the gauge coupling constant to the scalar field $\sigma$.

To obtain the classical solution in this theory, we impose the static,
axially-symmetric ansatz:
\begin{align}
\bm{A} &= \bm{e}_{\varphi} P(r)\,, \\
\sigma &= \rho(r) e^{i n\varphi}\,,
\end{align} 
where the integer $n$ is the winding number.
We used the circular cylindrical coordinates $r$, $\varphi$, and $z$.

We use the scale conversion $x \equiv gfr$, $\Tilde{P} \equiv g P$ and 
$\tilde{\rho}\equiv \rho/f$. Therefore the energy density of per unit
length of the $z$ axis becomes 
{\small\begin{equation}
{\cal E}=2 \pi f^2 \int^{\infty}_{0} dx \ x \left[ \dfrac{1}{2}
\left(\dfrac{\Tilde{P}'}{x} +\Tilde{\rho}^2 -1 \right)^2 +
\left( \Tilde{\rho}' + \dfrac{\Tilde{P}-n}{x} \Tilde{\rho} \right)^2 -
\dfrac{\Tilde{P}'}{x} \left( \Tilde{\rho}^2 - 1 \right) - 2\Tilde{\rho}
\Tilde{\rho}' \dfrac{\Tilde{P}-n}{x}  \right],
\end{equation}}%
where the prime ($'$) denotes the derivative with respect to $x$.
Asymptotic values are as follows:
$\tilde{P}(0)=0$, $\tilde{P}(\infty)=n$, $\tilde{\rho}(0)=0$ and
$\tilde{\rho}(\infty)=1$. We can write the following inequality for the
energy
\begin{equation}
{\cal E} \geq 2 \pi n f^2 \int^{\infty}_{0} \left( \Tilde{\rho}^2
\right)' dx=  2 \pi n f^2. 
\end{equation}
This lower bound on the energy is the Bogomolnyi bound and it is
saturated when $\Tilde{\rho}$ and $\Tilde{P}$ satisfy the following
equations
\begin{align}
\dfrac{\Tilde{P}'}{x} &= -\left( \Tilde{\rho}^2 -1\right), \\
\dfrac{\Tilde{\rho}'}{\Tilde{\rho}} &= - \dfrac{\Tilde{P}-n}{x}. 
\end{align}
These equations are the Bogomolnyi equations.

\section{Action and equation of motion with vortex ansatz}\label{detail}
In this Appendix, we show the details about the Bogomolnyi equations for
the vortex configuration. We take the axially symmetric ansatz:
\begin{equation}
\sigma_e=\rho_e(r) e^{in_e\varphi}\, ,\quad A_\varphi^v=P_v(r)\,.
\end{equation}
Then we find
\begin{equation}
{\cal D}_r\sigma_e=\rho'_e e^{in_e\varphi}\, ,\quad
{\cal D}_\varphi\sigma_e=i(n_e+(gP)_{t(e)}-(gP)_{o(e)})\rho_e
e^{in_e\varphi}\, ,
\end{equation}
where the prime denotes $\frac{d}{dr}$, the derivative with respect to
$r$, and
$(gP)_v=g_vP_v$. Thus the kinetic term of the scalar reads
\begin{equation}
|{\cal D}_i\sigma_e|^2=(\rho'_e)^2
+\frac{(n_e+(gP)_{t(e)}-(gP)_{o(e)})^2}{r^2}\rho_e^2\, ,
\end{equation}
while the Maxwell term becomes
\begin{equation}
\frac{1}{4}F_v^{ij}F_{ij}^v=\frac{1}{2}\frac{(P'_v)^2}{r^2}\,.
\end{equation}
The total action can be rewritten as
\begin{eqnarray}
{\cal E}&=&2\pi\int_0^\infty dr\,
r\left[\frac{1}{2}\sum_{v\in V}\frac{(P'_v)^2}{r^2}+
\sum_{e\in E}\left\{(\rho'_e)^2
+\frac{((E^TGP)_e-n_e)^2}{r^2}\rho_e^2\right\}\right. \nonumber \\
&~&\qquad\qquad\qquad+\left.\frac{1}{2}\sum_{e,e'\in
E}(\rho_e^2-\zeta_e)(E^TG^2E)_{ee'}(\rho_{e'}^2-\zeta_{e'})\right]\,,
\end{eqnarray}
and this is no other than the energy density per unit length
in the present static case. 

Varying this, we obtain the following equations of motion:
\begin{eqnarray}
\frac{(r\rho'_e)'}{r}&=&\frac{((E^TGP)_e-n_e)^2}{r^2}\rho_e+
\sum_{e,e'\in
E}\rho_e(E^TG^2E)_{ee'}(\rho_{e'}^2-\zeta_{e'})\,,\label{eom1}\\
\left(\frac{P'_v}{r}\right)'&=&2\sum_{e\in E}
\frac{((E^TGP)_e-n_e)}{r^2}\rho_e^2(E^TG)_{ev}\,.
\label{eom2}
\end{eqnarray}
These second-order simultaneous equations can be reduced to the
first-order
Bogomolnyi equations:
\begin{eqnarray}
\rho'_e&=&\mp\frac{(E^TGP)_e-n_e}{r}\rho_e\,,\label{mhb1}\\
\frac{P'_v}{r}&=&\mp\sum_{e\in
E}(\rho_{e}^2-\zeta_{e})(E^TG)_{ev}\,.
\label{mhb2}
\end{eqnarray}

\section{asymptotic profile of the vortex}\label{asy}
We investigate the asymptotic behavior of the solution of
(\ref{mhb1},\ref{mhb2}) in this Appendix.
To this purpose, first we introduce new variables $p_v(r)$ and $R_e(r)$:
\begin{equation}
P_v(r)=a_v-p_v(r)\,,\quad\rho_e=f_e-R_e(r)\,,
\end{equation}
where the constant $a_v$ satisfies
\begin{equation}
n_e=(E^TGa)_e\,.
\end{equation}

Next we prepare $p$-dimensional eigenvectors $x^{(a)}$ ($a=1,\cdots,p-1$)
for the
$(mass)^2$ mass matrix for vector fields satisfying
\begin{equation}
2(GEF)(GEF)^Tx^{(a)}=(m^{(a)})^2x^{(a)}\,,\quad
{\rm for~nonzero~modes}
\end{equation}
and $q$-dimensional eigenvectors $X^{(a)}$ for the $(mass)^2$
mass matrix for scalar fields satisfying
\begin{equation}
2(GEF)^T(GEF)X^{(a)}=(m^{(a)})^2X^{(a)}\,.
\end{equation}
Hereafter we restrict ourselves on the case with tree graphs treated in
the text. Thus $q=p-1$.
The zero mode satisfies
\begin{equation}
2(GEF)(GEF)^Tx^{(0)}=0\,.
\end{equation}

The relations of two sets of eigenvectors are
\begin{equation}
X^{(a)}=\frac{\sqrt{2}}{m^{(a)}}(GEF)^Tx^{(a)}\,,\quad
x^{(a)}=\frac{\sqrt{2}}{m^{(a)}}GEFX^{(a)}\,,\quad (a\ne 0)
\end{equation}
and we adopt the normalization convention:
\begin{equation}
x^{(a)T}x^{(a)}=X^{(a)T}X^{(a)}=1\,.
\end{equation}

Using the eigensystems,
we can expand the variables by eigenvectors as
\begin{equation}
p_v(r)=\sum_{(a)}p^{(a)}x^{(a)}_v\,,\quad
R_e(r)=\sum_{(a)}R^{(a)}X^{(a)}_e\,,
\end{equation}
Noticing 
$R_e(\infty)=0$ and $p_v(\infty)=0$,
the equations of motion (\ref{eom1},\ref{eom2}) becomes
at the asymptotic region, $r\rightarrow\infty$,
\begin{equation}
{R^{(a)}}''+\frac{1}{r}{R^{(a)}}'-(m^{(a)})^2{R^{(a)}}=0\,,
\end{equation}
\begin{equation}
{p^{(a)}}''-\frac{1}{r}{p^{(a)}}'-(m^{(a)})^2{p^{(a)}}=0\,,
\end{equation}
and the Bogomolnyi equations (\ref{mhb1},\ref{mhb2}) become
at the asymptotic region, $r\rightarrow\infty$,
\begin{equation}
{R^{(a)}}'=-\frac{1}{r}\frac{m^{(a)}}{\sqrt{2}}{p^{(a)}}\,,
\end{equation}
\begin{equation}
\frac{{{p^{(a)}}'}}{r}=-\sqrt{2}m^{(a)}{R^{(a)}}\,.
\end{equation}

The solution of the above equations is
\begin{equation}
R^{(a)}=C\,K_0(m^{(a)}r)\,,\quad p^{(a)}=\sqrt{2}C\,rK_1(m^{(a)}r)\,.
\end{equation}
This result can be derived by using the following formulas for the
modified Bessel function of the second type, such as $K_0(z)$ and 
$K_1(z)$;
\begin{equation}
K_0''(z)+\frac{1}{z}K_0'(z)-K_0(z)=0\,,\qquad
K_1''(z)+\frac{1}{z}K_1'(z)-\left(1+\frac{1}{z^2}\right)K_1(z)=0\,,
\end{equation}
\begin{equation}
(zK_1(z))''-\frac{1}{z}(zK_1(z))'-(zK_1(z))=0\,,
\end{equation}
\begin{equation}
K_0'(z)=-K_1(z)\,,\qquad (zK_1(z))'=-zK_0(z)\,,
\end{equation}
where the prime ($'$) means the derivative with respect to $z$.

More rough estimation can be done with the exponential function because
\begin{equation}
K_\nu(z)\approx\sqrt{\frac{\pi}{2z}}e^{-z}\,,\quad {\rm for~large~}z\,.
\end{equation}

\end{document}